diurnal9.tex
\documentstyle[12pt,epsfig]{article}
\begin{document}

\begin{titlepage}
\rightline{October 2011}
\vskip 2cm
\centerline{\Large \bf
Diurnal modulation due to self-interacting mirror}
\vskip 0.3cm
\centerline{\Large \bf 
 \& hidden sector dark matter}

\vskip 2.2cm
\centerline{R. Foot\footnote{
E-mail address: rfoot@unimelb.edu.au}}

\vskip 0.7cm
\centerline{\it ARC Centre of Excellence for Particle Physics at the Terascale,}
\centerline{\it School of Physics, University of Melbourne,}
\centerline{\it Victoria 3010 Australia}
\vskip 2cm
\noindent
Mirror and more generic hidden sector dark matter models can simultaneously explain 
the DAMA, CoGeNT and
CRESST-II dark matter signals consistently with the null results of the other experiments. This type of dark
matter can be captured by the Earth and shield detectors because it is self-interacting.
This effect will lead to a diurnal modulation in dark matter detectors. We estimate the size 
of this effect for dark matter detectors in various locations.
For a detector located in the northern hemisphere, this effect is expected to peak in April and 
can be detected for optimistic parameter choices.  
The diurnal variation is expected to be much larger for detectors located in the southern hemisphere. 
In particular, if the CoGeNT detector
were moved to e.g. Sierra Grande, Argentina then a $5 \sigma$ dark matter discovery would be possible in 
around 30 days of operation.

\end{titlepage}

\section{Introduction}

Mirror\cite{flv} and more generic hidden sector dark matter models 
provide a consistent explanation\cite{foot1,foot2,talk} of the DAMA annual modulation signal\cite{dama}, as well as the
low energy energy excess seen in the CoGeNT\cite{cogent} and CRESST-II\cite{cresst} experiments.
This explanation is consistent with the constraints from higher threshold experiments, such as CDMS\cite{cdms} and XENON100\cite{xenon100}
when reasonable systematic uncertainties in energy scale 
are considered\footnote{
There are also lower threshold analysis by the XENON10\cite{xenon10} and 
CDMS collaborations\cite{cdms8}.
However it has been argued\cite{collarguts} that neither analysis can
exclude light dark matter when systematic uncertainties are properly taken into account.}.

A distinctive feature of mirror and more generic hidden sector dark matter models is that the 
dark matter particles
are self-interacting. This means that dark matter captured by the Earth can effectively block
the halo dark matter wind. This effect will also lead to a diurnal modulation of detector dark matter 
signals. The purpose of this paper is to study this effect. Although we focus particular
attention on mirror and more generic hidden sector dark matter models with unbroken $U(1)'$
interaction, our study would be of relevance to many other self-interacting dark  
matter models\cite{otherdm}. 

Diurnal modulation of a dark matter signal can also occur due to elastic scattering of dark matter on
the constituent nuclei of the Earth\cite{collar}.
In that case, the effect is typically small unless the dark matter abundance is 
much less than expected.  We will show that the diurnal effect arising 
from the self-interactions of mirror/hidden sector particles can be much larger. In particular, for a detector
located in the southern hemisphere, the diurnal modulation effect is anticipated to be maximal, with 
a rate suppression varying between zero and near 100\% for a detector located in e.g. 
Sierra Grande, Argentina.
In fact, a dark matter experiment was conducted there in the 1990's\cite{arg}.
Unfortunately, the energy threshold was too high to detect
a significant
rate of dark matter interactions in that experiment.
However, CoGeNT's low threshold detector is highly portable
and could easily be moved from the Soudan laboratory to e.g. Sierra Grande. This would
allow the diurnal modulation to be detected with $5 \sigma$ C.L. in around 30 days of operation.

\section{Mirror \& hidden sector dark matter}

Mirror\cite{flv} and the more generic hidden sector dark matter models\cite{foot1,foot2008} assume the 
existence of a hidden sector which contains an 
unbroken $U(1)'$ gauge interaction
which is mixed with the standard $U(1)_Y$ via renormalizable kinetic mixing interaction:\cite{he,flv}
\begin{eqnarray}
{\cal L}_{mix} = \frac{\epsilon'}{2\cos\theta_w} F^{\mu \nu} F'_{\mu \nu}
\label{kine}
\end{eqnarray}
where $F_{\mu \nu}$ is the standard $U(1)_Y$ gauge boson 
field strength tensor, and $F'_{\mu \nu}$ is the field strength tensor for the hidden sector $U(1)'$. 
This interaction enables hidden sector $U(1)'$ 
charged particles (of charge $Qe$) to couple to
ordinary photons with electric charge $Q\epsilon' e \equiv \epsilon e$\cite{holdom}.
[In the case of mirror dark matter, $\epsilon' = \epsilon/\cos\theta_w$].
We consider the case where the hidden sector contains two (or more) stable $U(1)'$ charged 
dark matter particles, 
$F_1$ and $F_2$
with masses $m_1$ and $m_2$. 
Under the standard assumptions of a dark halo forming an
isothermal sphere, the condition of hydrostatic
equilibrium relates the halo temperature of the particles to
the galactic rotational velocity, $v_{rot} \sim 240\ {\rm km/s}$:
\begin{eqnarray}
T = {1 \over 2} \bar m v_{rot}^2
\end{eqnarray}
where $\bar m \equiv {n_{F_1} m_1 + n_{F_2} m_2 \over n_{F_1} + n_{F_2}}$
is the mean mass of the particles in the galactic halo.
We have assumed that the self interactions mediated by the unbroken $U(1)'$ gauge
interactions are sufficiently strong so that they thermalize the hidden sector 
particles, $F_1$ and $F_2$. The interaction length is typically much less than a parsec\cite{footz} and the
dark matter particles form a pressure-supported halo.
The dark matter particles are then described by a Maxwellian
distribution with $f_i(v) = exp(-E/T) = exp (-{1 \over 2} m_i v^2/T) = exp[-v^2/v_0^2 (i)]$
where
\begin{eqnarray}
v_0 (i) = v_{rot} \sqrt{{\bar m \over m_i}}\ . 
\label{sun}
\end{eqnarray}
With the assumptions that $m_2 >> m_1$ and that the abundance of $F_2$ is much less than $F_1$, we
have that $v_0^2 (F_2) \ll v_{rot}^2$. The narrow velocity dispersion (recall
$\sigma_v^2 = 3v_0^2/2$) can greatly reduce the rate
of dark matter interactions in higher threshold experiments such as CDMS\cite{cdms} and XENON100\cite{xenon100} whilst still explaining the
signals in the lower threshold DAMA and CoGeNT experiments.  

Mirror dark matter refers to the specific case where the hidden sector
is an exact copy of the standard model sector\cite{flv} (for a review
and more complete list of references see ref.\cite{review})
\footnote{
Note that successful big bang nucleosynthesis and successful
large scale structure requires effectively asymmetric initial
conditions in the early Universe, $T' \ll T$ and $n_{b'}/n_b \approx 5$. 
See ref.\cite{some} for further discussions.}.
In that case
a spectrum of dark matter particles of known masses are predicted: e$'$, H$'$, He$'$, O$'$, Fe$'$,... (with
$m_{e'} = m_e, m_{H'} = m_H,$ etc). 
The galactic halo is then presumed to be composed predominately of a spherically distributed 
self interacting mirror particle plasma comprising these particles\cite{sph}. 
Ordinary and mirror particles interact with each other via
kinetic mixing of the $U(1)_Y$ and its mirror counterpart. 
%Early Universe cosmology suggests\cite{pa}
%an upper limit $\epsilon \stackrel{<}{\sim} 10^{-9}$.

Both mirror and the more generic hidden sector dark matter models can
explain the direct detection experiments\cite{foot1,foot2,talk,footold1,footold}. 
These scenarios involve kinetic mixing induced
elastic (Rutherford) scattering of the dark matter particles off target nuclei.
In the mirror dark matter case, the DAMA/CoGeNT signals are assumed to arise from the
scattering of the dominant mirror metal component, $A'$, off target nuclei.
[The He$'$ and H$'$ components are too light to give a signal above the DAMA/CoGeNT energy threshold].
In the case of generic hidden sector models the role of $A'$ is played by $F_2$. 
Such elastic scattering can explain the 
normalization and energy dependence of the DAMA annual 
modulation amplitude and also the CoGeNT  and CRESST-II spectrum 
consistently with the null results of the other experiments, 
and yields a measurement of $\epsilon \sqrt{\xi_{A'}}$ and $m_{A'}/m_{F_2}$.
In the case of mirror dark matter the favoured parameter space is\cite{talk,foot2}:
\begin{eqnarray}
\epsilon \sqrt{\xi_{A'}} &\approx & {\rm few} \times 10^{-10}, \nonumber \\
\frac{m_{A'}}{m_p} &\approx & 16 - 56
\label{bla}
\end{eqnarray}
where $\xi_{A'} \equiv n_{A'}m_{A'}/(0.3 \ GeV/cm^3)$ is the halo mass fraction of the species
$A'$ and $m_p$ is the proton mass. 
In the case of the generic hidden sector model, $\epsilon$ is about an order of magnitude larger.
[The difference arises because in our notation the kinetic mixing induces an electric charge $\epsilon Z' e$ for
mirror dark matter and $\epsilon e$ for the generic hidden sector case].

As briefly discussed in Eq.(\ref{sun}) above, the velocity dispersion of the halo $A'/F_2$  
particles depends on the parameter $\bar m$. In the mirror model with isothermal halo
it can be estimated\cite{p3} to be around $1.1$ GeV
although it might vary
somewhat from this value. For  example, if the e$'$ have a 
lower temperature than the mirror nuclei in the halo then $\bar m$ can be somewhat larger, with an upper limit
of around $3$ GeV.
Although realistically it is unlikely that $\bar m$ could be this large in the mirror model, we consider this value
in our numerical work as an extremely optimistic case which might also be relevant to 
more generic hidden sector models where the
parameter $\bar m$ is less constrained.

\section{Distribution of mirror dark matter captured by the Earth and its shielding radius - $R_0$}

In the following sections we restrict our attention to the case of mirror dark matter for definiteness.
The more generic hidden sector scenario is completely analogous.
 
Mirror particles will occasionally be captured by the Earth and accumulate in the Earth's core.
Once a significant population of mirror particles have been accumulated, mirror
dark matter will be captured at a rate:
\begin{eqnarray}
{dN \over dt} \sim \pi R_0^2 f_{A'}
\end{eqnarray}
where $f_{A'} \approx v_{rot}\ \xi_{A'}' {0.3\ {\rm GeV/cm}^3 \over m_{A'}}$ 
is the flux of $A'$ mirror particles hitting the Earth.
Here $R_0$ is the maximum distance from the Earth's center for which dark matter
can be captured due to self interactions. We will show in a moment that
this distance is (currently) of order $4,000$ km for the example where $m_{A'} \approx 22 m_p$.
Thus we anticipate that around 
\begin{eqnarray}
N &\sim & \int  \pi R_0^2 f_{A'} dt \nonumber \\
&\sim & 10^{39} \left({\xi_{A'} \over 10^{-1}}\right)
\label{666}
\end{eqnarray}
$A'$ particles will be captured during the five billion year history of the Earth. This is many orders of magnitude 
within the geophysical limits\cite{slimit}.

A mirror particle, once trapped in the Earth, will lose energy rapidly due to interactions with 
the ordinary matter. The Rutherford cross section increases rapidly as the
relative velocity decreases, $d\sigma/d\Omega \propto 1/v^4$, and thus
one expects the captured mirror particles to rapidly thermalize with the ordinary matter.
Their distribution can then be obtained from the condition
of hydrostatic equilibrium:
\begin{eqnarray}
{dP \over dr} &=& -\rho_{A'} g \nonumber \\
{dn_{A'} \over dr} &=& - {n_{A'} \over T}\left[m_{A'} g +  {dT \over dr} \right]
\label{x}
\end{eqnarray}
where we have used $P = n_{A'}T$ and $\rho_{A'} = m_{A'} n_{A'}$. Note that 
we have implicitly assumed that captured mirror nuclei and mirror electrons 
combine into mirror atoms. The gravitational acceleration, $g$,
is given in terms of the mass density $\rho$,
\begin{eqnarray}
g = {G \over r^2}\int^r_0 \rho 4\pi r'^2 dr' \ .
\end{eqnarray}
Note that both $g$ and $T$ depend on $r$. We model  the 
Earth's density and temperature as given in figures 1a, 1b.
The boundaries in figure 1a correspond to the inner core, core, and mantle. In each of these regions
we use a linear approximation for the density adapted from the
Preliminary Reference Earth Model\cite{earthmodel}.
\vskip 0.2cm
\centerline{\epsfig{file=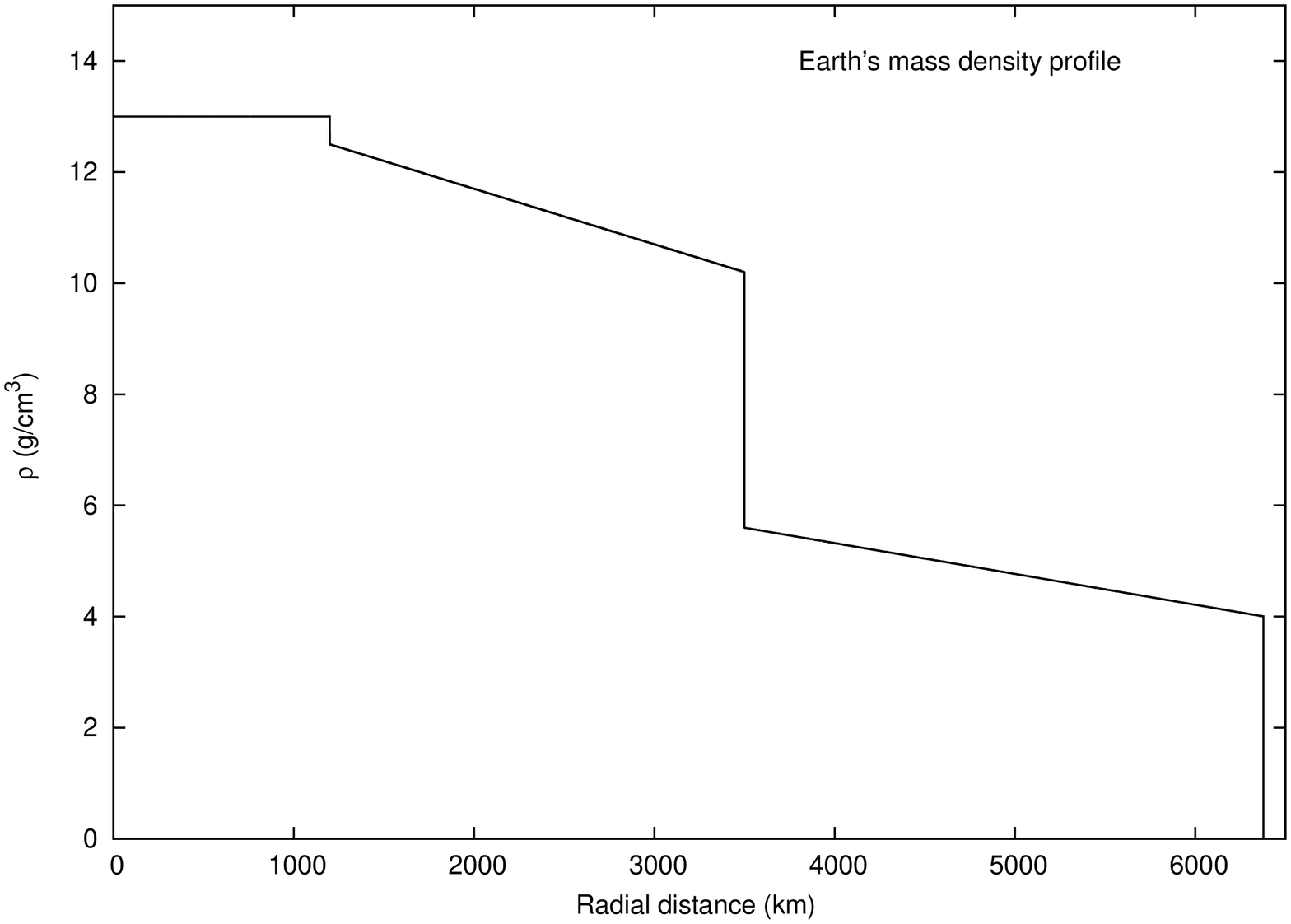,angle=270,width=13.0cm}}
\vskip 0.3cm
\noindent
{\small
Figure 1a: Earth's mass density [${\rm g/cm}^3$] versus radial distance. 
}

\centerline{\epsfig{file=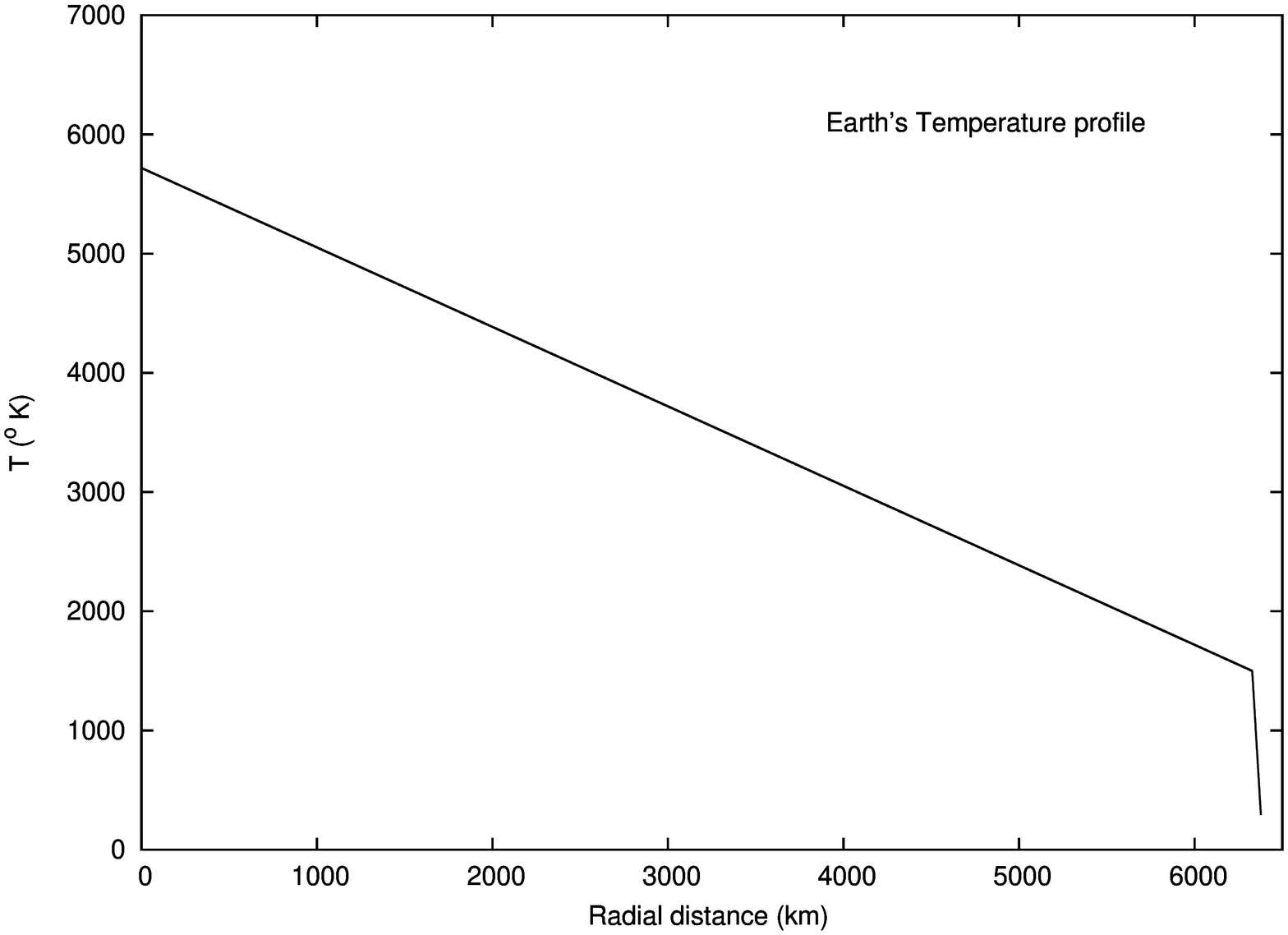,angle=270,width=13.0cm}}
\vskip 0.3cm
\noindent
{\small
Figure 1b: Earth's temperature profile [$^o K$]  versus radial distance.
%The dashed lines are the uncertainties in the temperature profile. 
}

\vskip 0.3cm
\noindent

Eq.(\ref{x}) can be numerically solved for $n_{A'} (r)$, 
taking into account the dependence of $g$ and $T$ on $r$.
The result is shown in figure 2. If the total number of accumulated $A'$ particles
is around $10^{39}$, as indicated in Eq.(\ref{666}), then we find that the central number density is of order 
$n_{A'}(0) \sim 10^{14}\ {\rm cm}^{-3}$.

\vskip 0.2cm
\centerline{\epsfig{file=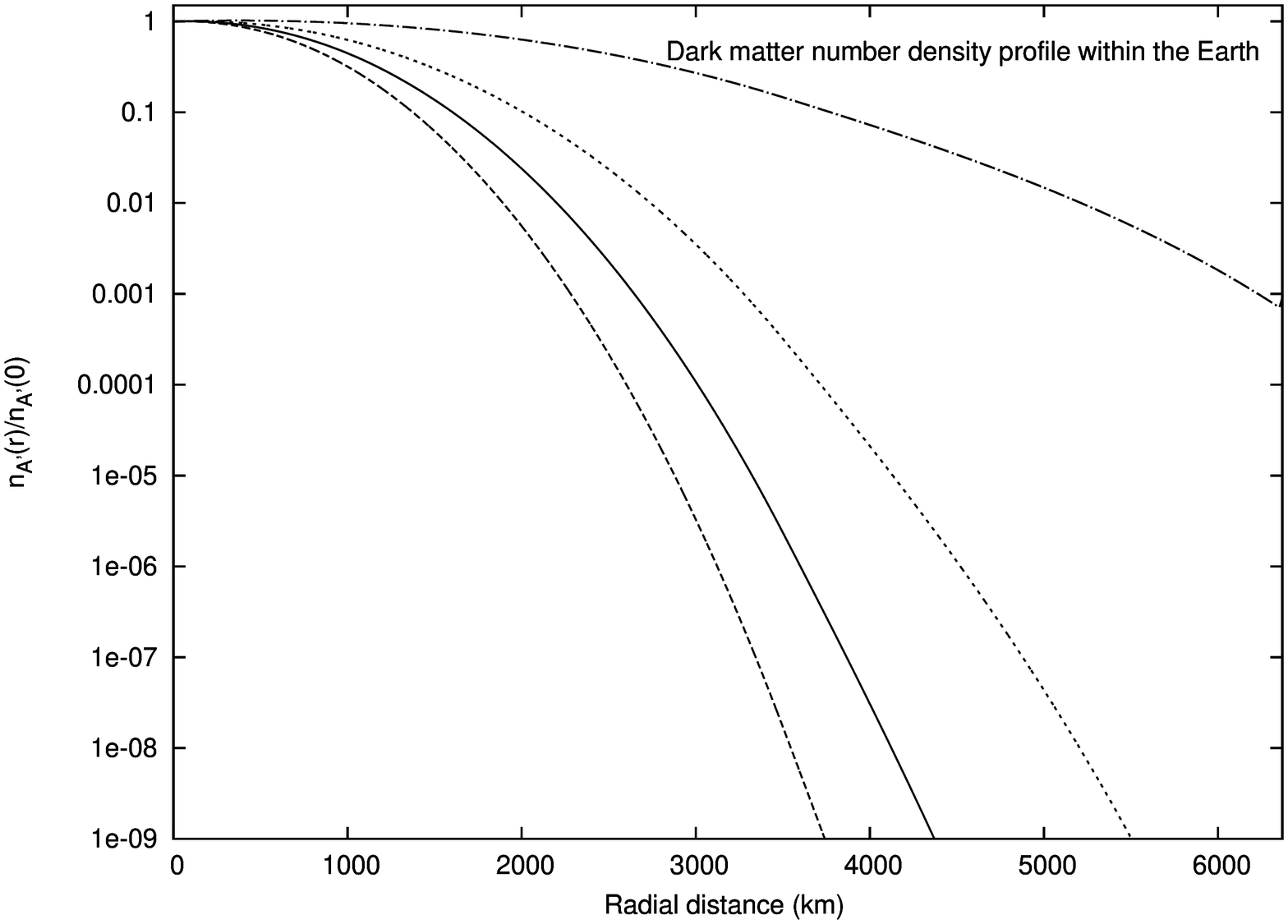,angle=270,width=15.0cm}}
\vskip 0.3cm
\noindent
{\small
Figure 2: The distribution within the Earth, $n_{A'}(r)/n_{A'}(0)$, of captured mirror 
particles of mass $m_{A'} = 22 m_p$ (solid line), $m_{A'} = 30 m_p$ (dashed line), $m_{A'} = 14m_p$ (dotted
line) and $m_{A'} = 4m_p$ (dashed-dotted line).
}
\vskip 1.2cm

Figure 2 suggests that
any captured He$'$ particles will quickly escape the Earth.
Indeed, the He$'$ distribution is quite spread out,
and particles in the tail of their Maxwellian velocity distribution
can have velocities much greater than the Earth's escape velocity. One can easily show\cite{footold1}
that e.g. He$'$ particles with velocities $\stackrel{>}{\sim} 20$ km/s can travel $\stackrel{>}{\sim} 100$ km
through ordinary matter without significant energy loss and escape the Earth's gravity (assuming
here $\epsilon \stackrel{<}{\sim} 10^{-9}$).
Heavier mirror particles on the other hand are much more tightly confined within the Earth
and have negligible probability of escaping Earth's gravity.

The captured mirror dark matter in the Earth can potentially shield a detector from halo dark matter
particles if they originate from a direction which takes them through the Earth.  Whether or not
a halo dark matter particle will get shielded will depend on how close its trajectory takes it
to the center of the Earth, where the density of Earth bound dark matter particles is greatest. 
Let us define $r_0$ as the distance of closest approach to the Earth's center 
of a given particle trajectory.
Let us now work out the maximum value of $r_0$, which we call $R_0$,
for which halo dark matter will be captured due to self interactions with Earth bound dark matter. 

When a dark matter particle, $A'$, from the halo passes through the dark matter accumulated in the 
Earth it will lose energy. The rate of energy loss is given by:
\begin{eqnarray}
{dE' \over dx} = - n_{A'}(x) \int^{E_R^{max}}_{E_R^{min}} {d\sigma \over dE_R} E_R dE_R \ ,
\label{d89}
\end{eqnarray}
where $x$ is the path of the dark matter particle.
If the density of Earth bound dark matter particles is great enough, the dark matter particle $A'$ will
lose nearly all of its energy from self-interactions and become trapped within the Earth.
The relevant cross-section for Rutherford scattering of a mirror nuclei with
incoming velocity $v$ on the mirror nuclei trapped within the Earth
(here both assumed to have atomic number $Z'$) and neglecting  
the form factors\footnote{We employ natural units where
$\hbar = c = 1$.} is given by
\begin{eqnarray}
{d\sigma \over dE_R} = {\lambda \over E_R^2 v^2}
\label{cs}
\end{eqnarray}
where 
\begin{eqnarray}
\lambda \equiv {2\pi Z'^4 \alpha^2 \over m_{A'}}   \ .
\end{eqnarray}
With this cross-section, Eq.(\ref{d89}) indicates that
dark matter particles will become trapped within the Earth if
their trajectories are such that:
\begin{eqnarray}
\int n_{A'}(x) dx &\stackrel{>}{\sim} & \  {{E'}^2_{in} \over 
m_{A'}\lambda \  log \left( {E_R^{max} \over E_R^{min}}\right) } \nonumber \\
& \stackrel{>}{\sim} & \left( {10 \over Z'} \right)^2 10^{16} \ {\rm cm}^{-2} \ ,
\label{991}
\end{eqnarray}
where $E'_{in} \sim {1 \over 2} m_{A'} v_{rot}^2$ is the initial energy of the dark matter
particle. The quantity $E_R^{max} = E'_{in}$ is the maximum kinematically allowed 
$E_R$, while $E_R^{min} \sim 1/(2r_b^2 m_{A'})$ ($r_b$ is the Bohr radius) is set by the scale
at which atomic screening becomes important.

Approximating the trajectories of dark matter particles by straight lines, and defining the 
co-ordinate $q$ as the distance along the trajectory with $r_0$ equal to the 
distance of closest approach to the Earth's center, we have:
\begin{eqnarray}
\int n_{A'}(x) dx = \int n_{A'}(r=\sqrt{r_0^2 + q^2}) dq \ .
\label{tied}
\end{eqnarray}
Eq.(\ref{991},\ref{tied}) can be numerically solved for $r_0$ assuming a given $m_{A'}$ value.
We find numerically that
dark matter particles will get captured by self interactions provided their trajectories 
satisfy
$r_0 \stackrel{<}{\sim} R_0$, with $R_0 \approx 4,000 \pm 700$ km, for $m_{A'}/m_p = 22 \pm 8$\footnote{
Note that even if heavy $\sim $ Fe$'$ component is responsible for the DAMA, CoGeNT and CRESST-II signals\cite{talk}, 
it is reasonable to expect lighter components do dominate the shielding within the Earth.}. 
Uncertainties in the
$\rho, T$ profile of the Earth give an 
(additional)
estimated systematic uncertainty of order $5-10\%$ in the determination of $R_0$ for a given
$m_{A'}$ value. Thus, $R_0$ could in principle be as large as $\approx 5,500 $ km.
This completes our estimation of the `shielding radius' $R_0$.

\section{Diurnal modulation effect due to shielding of the halo dark matter wind}

We are now ready to estimate the effect of the shielding of the halo mirror dark matter
due to its interactions with
mirror dark matter trapped within the Earth. In principle halo dark
matter interactions with ordinary matter
can also shield or at least reduce the energy of halo dark matter particles\cite{collar}. 
However, for the case of mirror dark matter the latter effect is expected to be relatively small\cite{footold1}
especially if the kinetic mixing parameter $\epsilon$ satisfies the cosmology bound\cite{disf} 
$\epsilon \stackrel{<}{\sim} 10^{-9}$.

The direction of the Earth's motion through the halo, subtends an (average) angle $\approx 43^o$ with 
respect to the Earth's spin axis.
There is a small annual modulation 
in this angle due to the Earth's rotation around the sun, which can be important and will be
considered in the following section. 
Another relevant angle is the angle between the direction of the Earth's motion through
the halo and the normal vector to the Earth's surface at the detector's location. This angle is denoted by $\psi$
[Note that in ref.\cite{collar} this angle was denoted as $\theta$]. 
The angle $\psi$, depends on the detector location and also the time of day, $t$:
\begin{eqnarray}
\cos \psi = \cos \theta_{latitude} \sin \omega t \sin 43^o \pm \sin \theta_{latitude} \cos 43^o
\label{42}
\end{eqnarray}
where $\omega = 2\pi/T_d$ with $T_d = 1$ sidereal day (23.934 hours).
In the above equation, the $+$ [$-$] sign is relevant for a northern [southern] hemisphere detector, 
where $\theta_{latitude}$ is the north [south] latitude.

Dark matter particles arriving at a detector originate predominately from a cone with axis in the direction
of the Earth's motion through the halo and with the detector at its apex. 
The particles within the cone which pass through
a distance from the Earth's center of $r_0 \stackrel{<}{\sim} R_0$ ($R_0 \approx 4,000$ km) 
will be shielded from the detector, while
particles with $r_0 \stackrel{>}{\sim} R_0$  will arrive unhindered by self-interactions.
Because the direction of the cone of dark matter particles periodically changes due to the Earth's
rotation, the proportion of dark matter particles shielded also modulates. Our task now is
to estimate this effect for detectors in various locations.

The differential interaction rate is given by:
\begin{eqnarray}
{dR \over dE_R} &=& 
 N_T n_{A'} \int {d\sigma \over dE_R} {f_{A'}({\bf{v}},{\bf{v}}_E) \over
k} |{\bf{v}}|
d^3v \nonumber \\
&=& N_T n_{A'}
{\lambda \over E_R^2 } \int^{\infty}_{|{\bf{v}}| > v_{min}
(E_R)} {f_{A'}({\bf{v}},{\bf{v}}_E) \over k|{\bf{v}}|} d^3 v 
\label{55}
\end{eqnarray}
where $N_T$ is the number of target atoms per kg of detector,
$k = (\pi v_0^2 [A'])^{3/2}$ is the Maxwellian distribution
normalization factor and
$n_{A'} = \rho_{dm} \xi_{A'}/m_{A'}$ is the number density of the halo mirror
nuclei $A'$ at the Earth's
location (we take $\rho_{dm} = 0.3 \ {\rm GeV/cm}^3$). [This should not be
confused with the number density of trapped mirror nuclei in the Earth, despite
the similarity in the notation!].
Here ${\bf{v}}$ is the velocity of the halo particles relative to the
Earth and ${\bf{v}}_E$ is the
velocity of the Earth relative to the galactic halo.
The halo distribution function in the reference frame of the Earth is given by, 
$f_{A'} ({\bf{v}},{\bf{v}}_E)/k = (\pi v_0^2[A'])^{-3/2} exp(-({\bf{v}}
+ {\bf{v}}_E)^2/v_0^2[A'])$. 
Note that the lower velocity limit,
$v_{min} (E_R)$, 
is given by the kinematic relation:
\begin{eqnarray}
v_{min} &=& \sqrt{ {(m_A + m_{A'})^2 E_R \over 2 m_A m^2_{A'} }}
\ .
\label{v}
\end{eqnarray}

The interaction rate depends on the velocity integral in Eq.(\ref{55}),
\begin{eqnarray}
I \equiv \int^{\infty}_{|{\bf{v}}| > v_{min}
(E_R)} {f_{A'}({\bf{v}},{\bf{v}}_E) \over k|{\bf{v}}|}  d^3 v \ .
\label{55x}
\end{eqnarray}
This integral can be modified to incorporate the effect of shielding of the halo dark matter
by the Earth bound mirror matter, by multiplying the integrand  by the quantity $g(\theta, \phi,\psi)$,
where 
\begin{eqnarray}
g (\theta, \phi,\psi) &=& 0   \  {\rm if}\ d_{min} < R_0 \nonumber \\
&=& 1 \ {\rm if}\ d_{min} > R_0 \ .
\end{eqnarray}
Here $d_{min}$ is the distance of closest approach to the center of the Earth experienced by
the particle trajectories. We show in the 
appendix that $d_{min}$ is given by:
\begin{eqnarray}
d^2_{min} (\theta,\phi,\psi) &=& R_E^2 [1 - f^2(\theta,\phi,\psi)] \ {\rm if}\ f(\theta,\phi,\psi) > 0 \nonumber \\
          &=& R_E^2\ {\rm if}\ f(\theta,\phi,\psi) < 0
\label{18}
\end{eqnarray}
where 
\begin{eqnarray}
f(\theta,\phi,\psi) =\sin\theta \sin\psi \sin\phi - \cos\theta \cos\psi
\end{eqnarray}  
and $R_E \simeq 6378$ km is the radius of the Earth.
Thus, the shielding effects can be incorporated by replacing $I$, defined in Eq.(\ref{55x}), with:
\begin{eqnarray}
I[\psi (t)] &\equiv & \int^{\infty}_{|{\bf{v}}| > v_{min}
(E_R)} {f_{A'}({\bf{v}},{\bf{v}}_E)  \over k|{\bf{v}}|} g(\theta, \phi, \psi) d^3 v \nonumber \\
&\equiv & \int_{-1}^{1} \int_{0}^{2\pi}  \int^{\infty}_{|{\bf{v}}| > v_{min}
(E_R)} {f_{A'}({\bf{v}},{\bf{v}}_E) \over k} g(\theta, \phi, \psi) |{\bf{v}}| 
d|{\bf{v}}| d\phi d\cos\theta \ . 
\label{55x2}
\end{eqnarray}

We are now in a position to evaluate the effect of shielding of mirror dark matter expected for 
experiments at
various locations. We compute the quantity  
\begin{eqnarray}
{\cal R} \equiv 100 \times \left(1 - {I[\psi (t)] \over I} \right)
\end{eqnarray}
which gives the percentage rate reduction
in the interaction rate due to shielding of dark matter in the core of the Earth.
In figure 3a we give results for ${\cal R}$, with $E_R = 6.7 \ keV_{NR}$ for Na target,
relevant for  the DAMA set up at Gran Sasso with north latitude = $42.5^o$.
We consider two reference sets of parameters, the 
first one  assumes the `standard' parameters  $R_0 = 4,000$ km and $\bar m = 1.1 $ GeV. 
The second possibility assumes
`optimistic' values $R_0 = 5,500$ km and $\bar m = 3.0$ GeV. 
Figure 3a shows that even with the optimistic parameters, the diurnal signal modulation
is expected to be relatively small for DAMA/Libra at Gran Sasso. Thus the lack of a diurnal modulation
in the DAMA/NaI experiment\cite{de} does not significantly constrain any parameter space.
Currently DAMA/Libra is running with a lower energy threshold which will make them
more sensitive to the diurnal effect simply because the signal rate is much higher at lower
energies, $dR/dE_R \propto 1/E_R^2$. 
Thus, for optimistic parameters DAMA/Libra might possibly detect a small
diurnal modulation.

The CoGeNT Ge detector has been operating in the Soudan underground laboratory, which 
has north latitude of $48^o$. This is a little further north than that of Gran Sasso, so the
diurnal modulation effect expected at Soudan is smaller than that for DAMA/Libra.
However, for the CDEX/Texono Ge detectors\cite{wong} located at Jin-Ping underground laboratory at
north latitude $28^o$, the effects are expected to be larger.
In figure 3b we give the results for these detectors  
at $E_R = 2 \ keV_{NR}$.

\vskip 0.2cm
\centerline{\epsfig{file=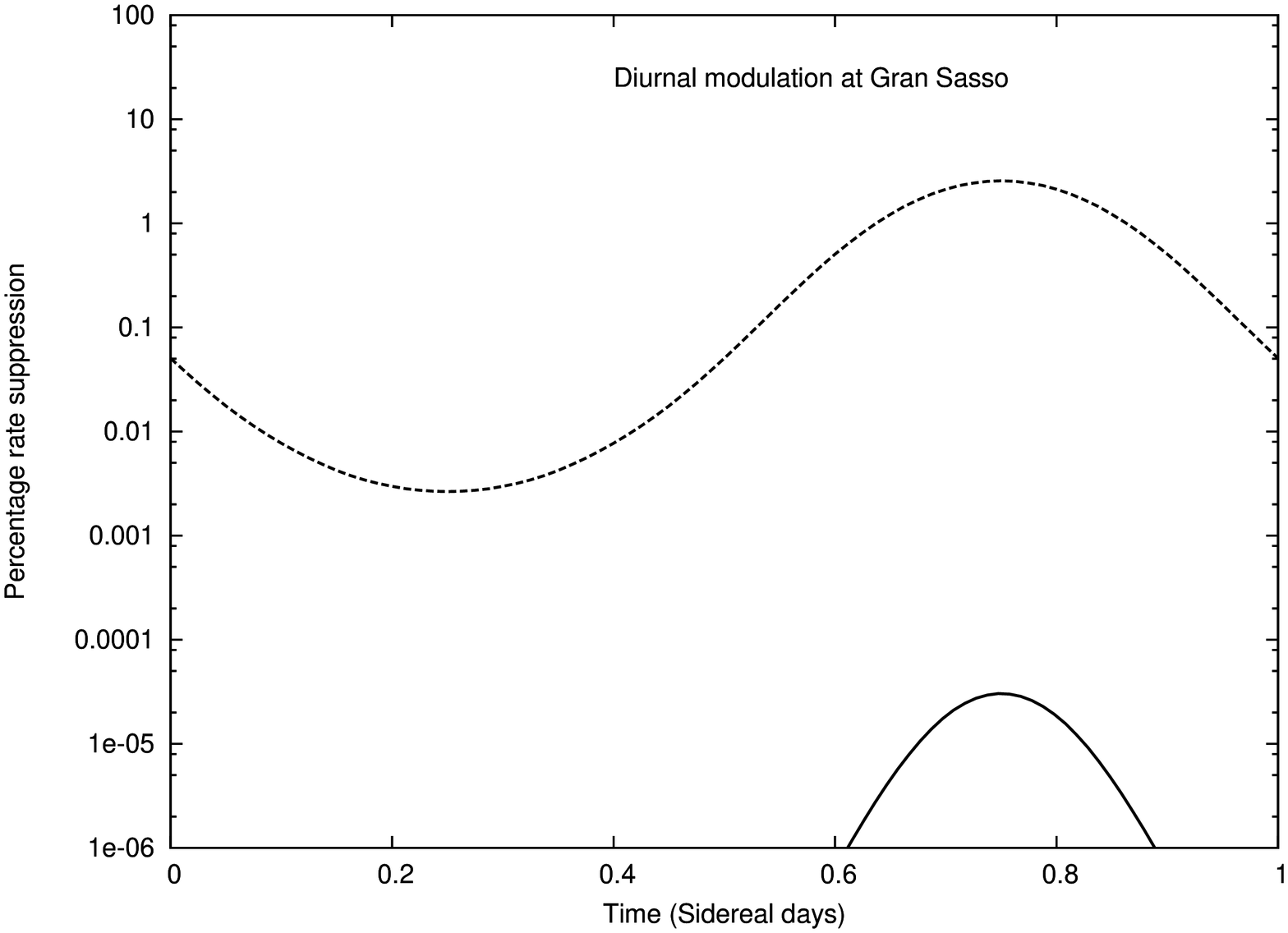,angle=270,width=15.0cm}}
\vskip 0.3cm
\noindent
{\small
Figure 3a: Percentage rate suppression due to the shielding of 
dark matter in the Earth's core versus time, for a Na detector
at Gran Sasso.  
A reference energy of $6.7\ keV_{NR}$ 
has been assumed. Both the captured and interacting dark matter particles 
are assumed to have the 
mass $m_{A'} = 22 m_p$. 
The solid line assumes `standard' parameters where the shielding radius is
taken to be $R_0 = 4,000$ km and the halo velocity dispersion is given in Eq.(\ref{sun}) 
assuming $\bar m = 1.1$ GeV.
The dashed line is for the more optimistic case considered where
$R_0 = 5,500$ km and $\bar m = 3.0$ GeV.
}

\vskip 0.2cm
\centerline{\epsfig{file=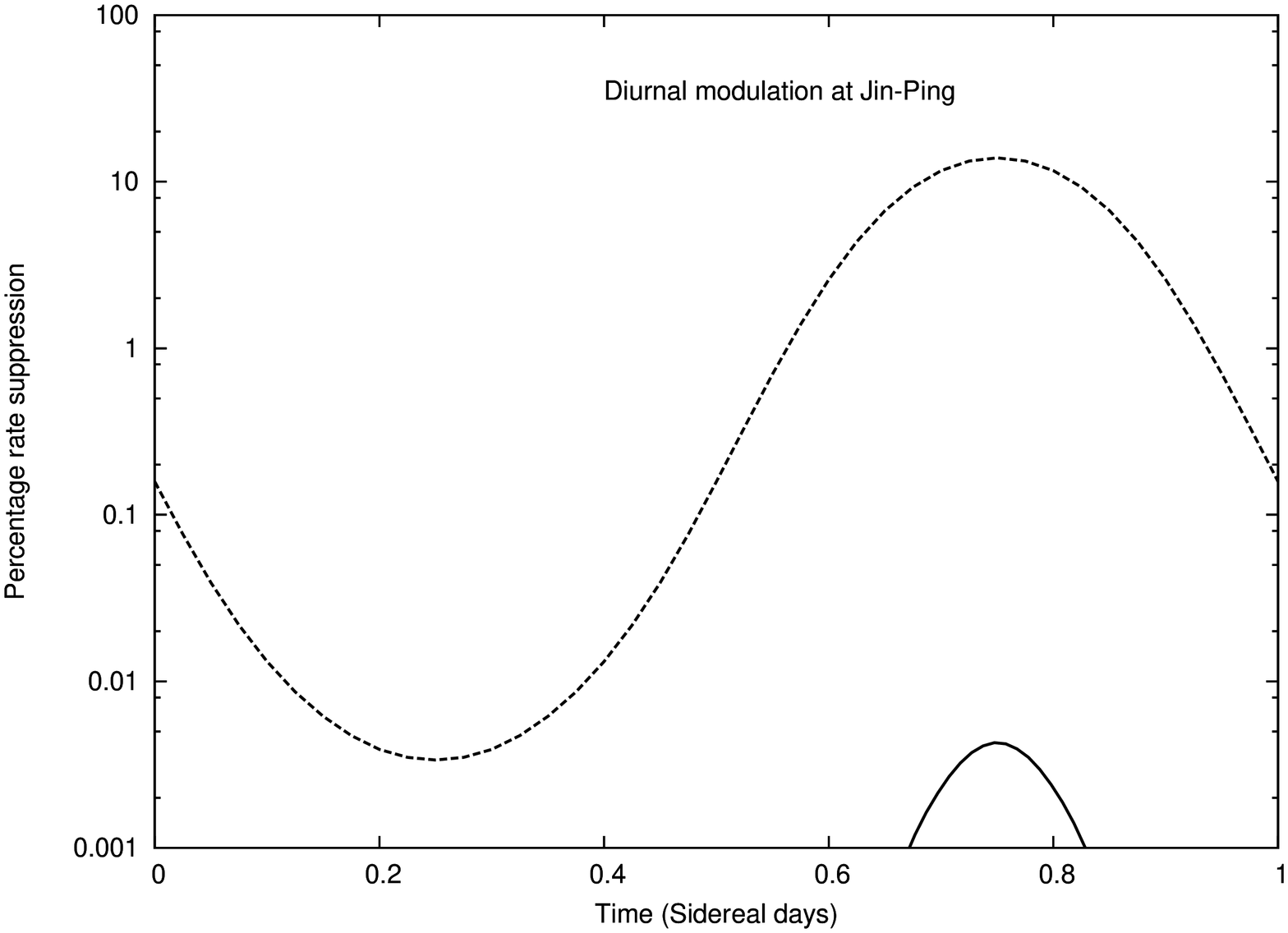,angle=270,width=15.0cm}}
\vskip 0.3cm
\noindent
{\small
Figure 3b: Same as figure 3a except that the Na detector at Gran Sasso is changed to
CDEX/Texono Ge detector at Jin-Ping underground laboratory (a reference energy
of $2 \ keV_{NR}$ is assumed).}

\vskip 0.8cm

The effects in existing detectors in the northern hemisphere are not very large, 
although there is some possibility of
observing the effect with optimistic parameters, especially at the Jin-Ping laboratory. 
Of course, the 
reason that the shielding effect is not so large is simply because all of the existing detectors are located
in the northern hemisphere. The effect becomes dramatically bigger for 
a detector located in the southern hemisphere because the halo dark matter particles then
have a much larger probability of passing through the center of the Earth
%\footnote{footnote about detectorat south pole xxxxx}. 
We illustrate this in figure 3c
which gives results for the CoGeNT detector relocated to the Sierra Grande laboratory
in Argentina, with south latitude $41.6^o$. 
As the figure shows, the variation ranges from near zero to 100 \%. At Soudan, the CoGeNT
spectrum is consistent with around 3 dark matter signal events per day per $0.3 \ kg$. 
We estimate then that the diurnal effect at Sierra Grande would be so large that just 30 days 
of data would be sufficient to yield a $5\sigma$ discovery of dark matter with CoGeNT's $\sim 0.3$ kg 
detector.

\vskip 0.2cm
\centerline{\epsfig{file=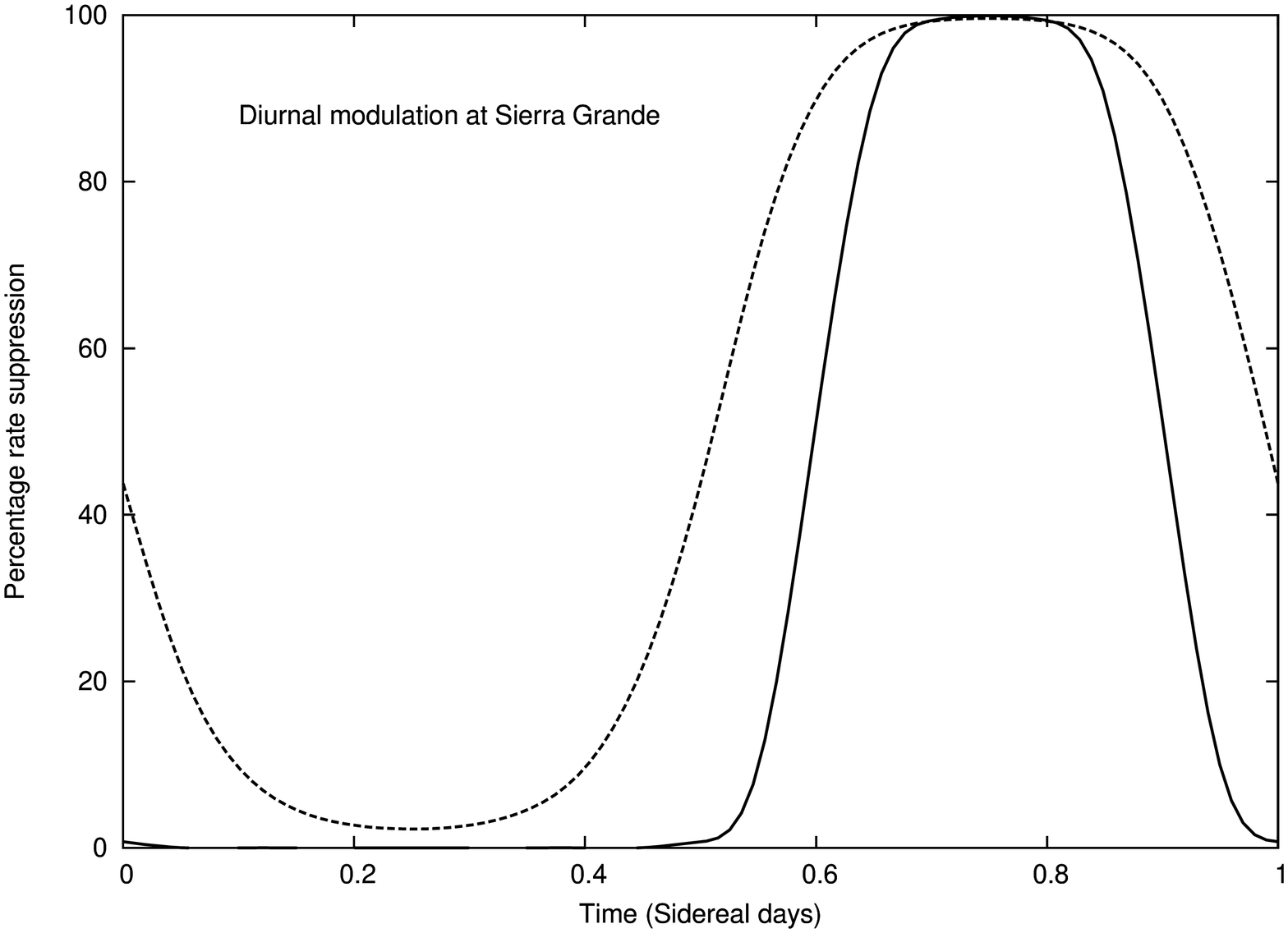,angle=270,width=15.0cm}}
\vskip 0.3cm
\noindent
{\small
Figure 3c: Same as figure 3a except that the Na detector at Gran Sasso is changed to
CoGeNT Ge detector relocated to Sierra Grande, Argentina.}

\vskip 1.1cm

Finally, we comment on the proposal to place a NaI detector at the South Pole\cite{sp}.
Interestingly, the South Pole is the only point in the southern hemisphere 
where the diurnal modulation effect is absent. For a detector at the south pole
we find a constant $32\%$ [$85\%$] rate suppression for the
standard [optimistic] parameter values.

\section{Annual variation of the diurnal modulation}

The proportion of the Earth which shields dark matter particles has both a diurnal variation and an
annual variation. 
The annual variation, which we have hitherto neglected, 
arises because of the variation of the direction of the Earth's motion
through the halo, due to the Earth's motion around the sun.
In this section we examine this effect.

The direction of the Earth's motion through the halo, subtends an angle $\theta_1 (t)$ 
with respect to the Earth's spin axis.
On average this angle is approximately $43^o$, but varies during the year.
To evaluate this angle at a  given time, $t$, define a right-angled co-ordinate system with 
the z-axis in the direction of the normal to the ecliptic plane, x-axis in the direction
of the sun relative to the Earth  
and y-axis in the direction of the Earth's motion around the sun.
In this co-ordinate system, for an observer at rest with respect to the halo, 
the Earth's velocity vector is
\begin{eqnarray}
{\bf v}_E &=& v_\odot [-\sin 30^o \cos 2\pi{(t - t_1) \over T}, \ -\sin 30^o \sin 2\pi {(t - t_1)\over T}, \ \cos 30^o] + 
v_\oplus [0, 1, 0]
\nonumber \\
&=& v_\odot [-\sin 30^o \cos 2\pi {(t - t_1)\over T}, \ -\sin 30^o \sin 2\pi {(t - t_1)\over T} + y, \ \cos 30^o]
\end{eqnarray} 
where $t_1 = 152.5 \ {\rm days} + 0.25 \ {\rm years} \simeq 244$ days, $T = 1$ year and $y = v_\oplus/v_\odot \approx
30/240 \approx 0.125$.
The angle, $30^o$ is the angle between the normal of the ecliptic plane and 
the sun's direction of motion through
the halo. The
Earth's spin axis has the direction:
\begin{eqnarray}
{\bf L}_{spin} = [\sin 23.5^o \cos 2\pi {(t-t_2) \over T}, \ \sin 23.5^o \sin 2\pi {(t-t_2) \over T}, 
\ \cos 23.5^o]
\end{eqnarray}
where $t_2 \simeq 172$ days (northern summer solstice).
The angle $23.5^o$ is the tilt of the Earth's spin axis relative to the normal of the ecliptic plane.
The angle $\theta_1 (t)$, is then given by 
\begin{eqnarray}
\cos\theta_1 &=& {{\bf v}_E \cdot {\bf L}_{spin} \over |{\bf v}_E|\ |{\bf L}_{spin}|} \nonumber \\
 & \simeq  & \cos \bar \theta_1 + y\left[ \cos \bar \theta_1 \sin 30^o \sin 2\pi {(t-t_1) \over T} 
+ \sin 23.5^o \sin 2\pi {(t-t_2) \over T} \right]
\end{eqnarray}
where
\begin{eqnarray}
\cos \bar \theta_1 = \cos 30^o \cos 23.5^o - \sin 30^o \sin 23.5^o \cos 2\pi{(t_1 - t_2) \over T}
\end{eqnarray}
Thus, we see that the angle $\theta_1 (t)$ is not actually a constant, but has a small
variation due to the earth's motion around the sun ($y \neq 0$).

\vskip 0.2cm
\centerline{\epsfig{file=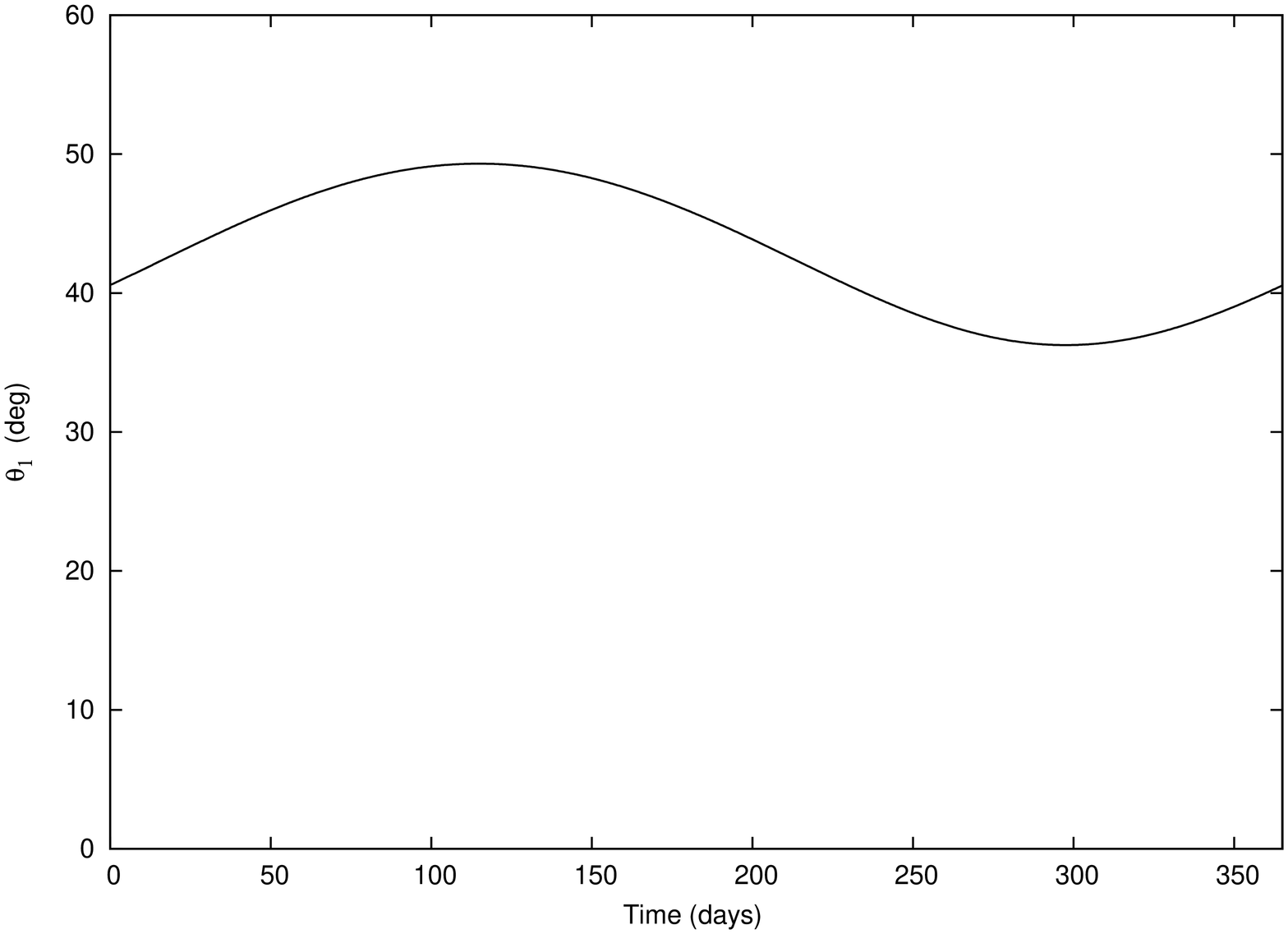,angle=270,width=15.0cm}}
\vskip 0.3cm
\noindent
{\small
Figure 4: The angle, $\theta_1 (t)$ versus time. [This is the angle 
between the Earth's spin axis and the Earth's direction of motion 
through the halo].}

\vskip 0.8cm

In figure 4, we plot $\theta_1 (t)$ over its period of one year (where $t=0$ corresponds to January 1).
As the figure shows, we find that  $\theta_1$ has a maximum of $49.3^o$ (April 25) and minimum of 
$36.3^o$ (October 25). This variation is relatively unimportant for a detector in the
southern hemisphere\footnote{For the special case of a detector at the South Pole, 
there is no diurnal signal as 
noted earlier,
but there is a rate suppression which annually modulates slightly due to the variation of $\theta_1 (t)$.}. 
For a detector in the northern hemisphere we expect
the largest diurnal signal to occur when $\theta_1$ is a maximum and thus we anticipate that the largest 
diurnal signal will occur in April.  The effect of the $\theta_1$ variation can be estimated by 
repeating the analysis of the previous section,
with $43^o$ in Eq.(\ref{42}) replaced with $49.3$ and $36.3$ [in general $43^o$ is 
replaced by $\theta_1 (t)$ in Eq.(\ref{42})]. In figure 5a we plot the percentage rate 
suppression for the optimistic
parameter choice for a detector located at Gran Sasso. The figure shows that the 
variation of the diurnal signal during
the year is significant. Figure 5b is the corresponding results for a detector located at Jin-Ping
Underground laboratory.
Clearly the March - May quater provides the best window to detect the diurnal signal for a northern hemisphere
detector. The CDEX experiment in Jin-Ping, currently running with a 1 kg Ge PC detector,
with plans to upgrade to $> 10$ kg has an excellent opportunity to observe the diurnal signal.
The 250 kg DAMA/Libra detector, currently running with lower threshold, also has a very
good chance to see such a signal in the March-May quater.

\vskip 0.2cm
\centerline{\epsfig{file=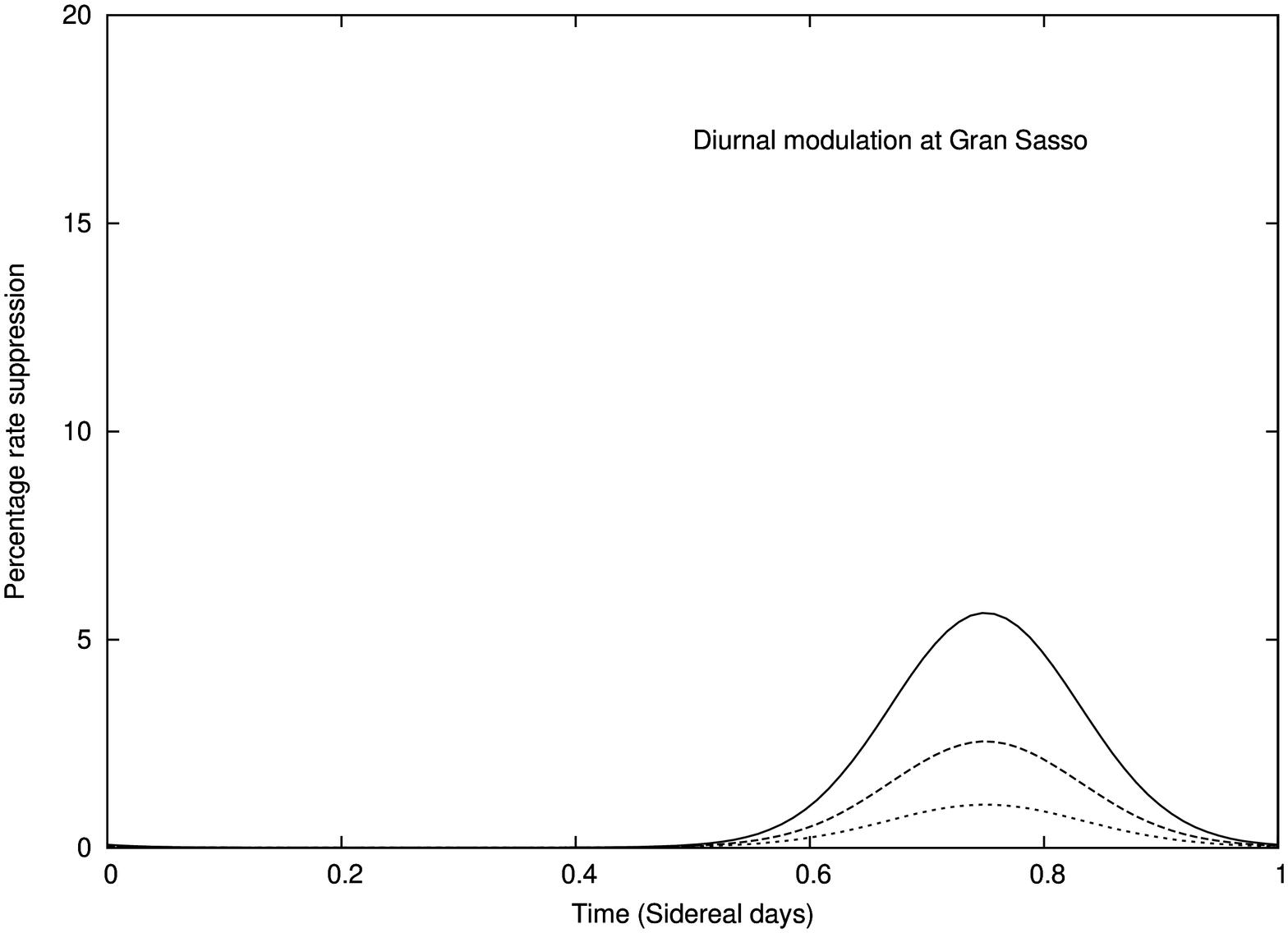,angle=270,width=15.0cm}}
\vskip 0.3cm
\noindent
{\small
Figure 5a: Percentage rate suppression due to the shielding of 
dark matter in the Earth's core versus time, for a Na detector
at Gran Sasso.  
A reference energy of $6.7\ keV_{NR}$ 
has been assumed. Both the captured and interacting dark matter particles 
are taken to have the 
reference mass $m_{A'} = 22 m_p$. 
We have assumed the optimistic case with shielding radius of 
$R_0 = 5,500$ km and the halo velocity dispersion is given in Eq.(\ref{sun}) 
with $\bar m = 3.0$ GeV.
The solid line assumes $\theta_1 = 49.3^o$ (April 25),
the dashed line assumes $\theta_1 = 43^o$, (Yearly average) and the dotted line assumes
$\theta_1 = 36.3^o$ (October 25).
}

\vskip 0.2cm
\centerline{\epsfig{file=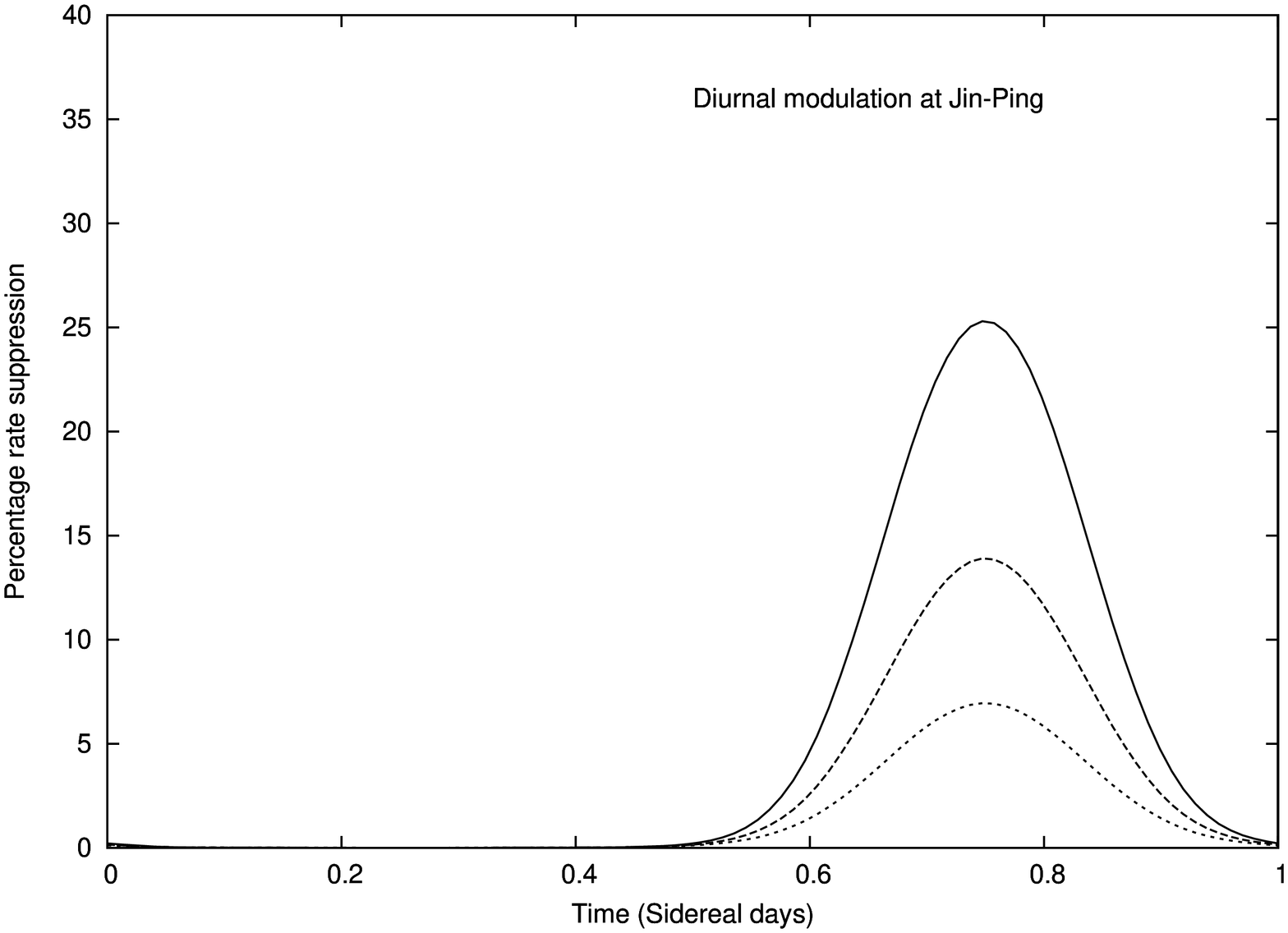,angle=270,width=15.0cm}}
\vskip 0.3cm
\noindent
{\small
Figure 5b: Same as figure 5a except that the Na detector at Gran Sasso is changed to
CDEX/Texono Ge detector at Jin-Ping underground laboratory (a reference energy
of $2 \ keV_{NR}$ is assumed).}

\vskip 0.8cm

\section{Conclusion}

Mirror and more generic hidden sector dark matter models can simultaneously explain 
the DAMA, CoGeNT and
CRESST-II dark matter signals consistently with the null results of the other experiments. 
This type of dark
matter can be captured by the Earth and shield detectors because it is self-interacting.
This effect will lead to a diurnal modulation in dark matter detectors. We have estimated the size 
of this effect for dark matter detectors in various locations.

Current dark matter detectors in the northern hemisphere are not so sensitive to the diurnal modulation
although there is an exciting possibility that it can be detected at Jin-Ping and even in DAMA/Libra
for optimistic parameter choices. We have also found that the diurnal signal for detectors located 
in the northern hemisphere varies during the year, peaking in April.

The situation changes dramatically for detectors located in the southern hemisphere where
the dark matter wind passes directly through the core of the Earth for part of the day.
In particular, if the CoGeNT detector
were moved to e.g. Sierra Grande, Argentina then a $5 \sigma$ dark matter discovery would be possible in 
around 30 days of operation.

\vskip 0.5cm
\noindent
{\Large \bf  Appendix: Calculation of $d_{min}$}

\vskip 0.5cm

In this appendix we derive Eq.(\ref{18}). Recall dark matter particles reaching a detector
originate predominately from a cone centered along the Earth's direction of motion through the halo.
Recall also that $\psi$ is the angle between the normal vector to the Earth's surface 
at the detector's location and the axis of this cone.
We assume that particles
form straight line trajectories. Let us introduce a co-ordinate system with
the detector at the origin. We use polar co-ordinates with the $z-axis$ aligned with the direction
of the Earth's motion through the halo.
The points along the particle trajectory are given by 
$(r\sin\theta \sin\phi, r\sin\theta\cos\phi, r\cos\theta)$. In this co-ordinate system (suitably rotated in
the azimuthal direction)
the location of Earth's center is 
$(R_E \sin \psi, 0, -R_E \cos\psi)$, where $R_E \simeq 6378$ km is the Earth's radius. 
The distance between the Earth's
center and a point along the particle trajectory is then given by
\begin{eqnarray}
d^2 = R_E^2 + r^2 + 2R_E r (\cos\theta \cos\psi - \sin\theta \sin\psi \sin\phi)\ .
\label{a11}
\end{eqnarray}
We need to determine $d_{min}$ which is the distance of closest approach to the 
Earth's center as we vary $r$ keeping $\theta, \phi$ and $\psi$ fixed. 
The value of $r$ where this occurs can be determined by minimizing
$d^2$ with respect to $r$, i.e. from $\partial d^2/\partial r = 0$. 
Evaluating this value of $r$ and substituting it
back into Eq.(\ref{a11}) we obtain Eq.(\ref{18}) as the solution.
This completes the outline of the derivation of Eq.(\ref{18}).

\vskip 1cm
\noindent
{\large Acknowledgments}

\vskip 0.2cm
\noindent
This work was supported by the Australian Research Council.


\begin{thebibliography}{999}

\bibitem{flv}
R. Foot, H. Lew and R. R. Volkas, Phys. Lett. B272, 67 (1991);
Mod. Phys. Lett. A7, 2567 (1992).

\bibitem{foot1}
R. Foot, Phys. Lett. B703, 7 (2011) [arXiv: 1106.2688]; 
Phys. Lett. B692: 65 (2010) [arXiv: 1004.1424].

\bibitem{foot2}
R. Foot, Phys. Rev. D82: 095001 (2010) [arXiv: 1008.0685].

\bibitem{talk}
R. Foot, arXiv: 1203.2387.

\bibitem{dama}
R. Bernabei {\it et al}. (DAMA Collaboration), 
Riv. Nuovo Cimento. 26, 1 (2003) [astro-ph/0307403]; Int. J. Mod.
Phys. D13, 2127 (2004); Phys. Lett. B480, 23 (2000);
Eur. Phys. J. C56: 333 (2008) [arXiv:0804.2741]; Eur. Phys. J. C67, 39 (2010) [arXiv: 1002.1028].

\bibitem{cogent}  
C. E. Aalseth {\it et al.} (CoGeNT Collaboration),  Phys. Rev. Lett. 106: 131301 (2011)
[arXiv:1002.4703]; Phys.Rev.Lett. 107 (2011) 141301 [arXiv: 1106.0650].

\bibitem{cresst}
G. Angloher {\it et al.}, (CRESST Collaboration),
arXiv:1109.0702.


\bibitem{cdms}
Z. Ahmed {\it et al} (CDMS Collaboration), Science 327: 1619 (2010) [arXiv: 0912.3592].

\bibitem{xenon100}
E.~Aprile {\it et al.}  (XENON100 Collaboration),
%``Dark Matter Results from 100 Live Days of XENON100 Data,''
Phys.\ Rev.\ Lett.\   107, 131302 (2011)
[arXiv:1104.2549].
%%CITATION = PRLTA,107,131302;%%

\bibitem{xenon10}
J.~Angle {\it et al.}  [XENON10 Collaboration],
%``A search for light dark matter in XENON10 data,''
Phys.\ Rev.\ Lett.\  107,  051301 (2011)
[arXiv:1104.3088].
%%CITATION = PRLTA,107,051301;%%

\bibitem{cdms8}
Z.~Ahmed {\it et al.}  [CDMS-II Collaboration],
%``Results from a Low-Energy Analysis of the CDMS II Germanium Data,''
Phys.\ Rev.\ Lett.\  106, 131302 (2011)
[arXiv:1011.2482].
%%CITATION = PRLTA,106,131302;%%


\bibitem{collarguts}
J.~I.~Collar,
%``A comparison between the low-energy spectra from CoGeNT and CDMS,''
arXiv:1103.3481 ;
%%CITATION = ARXIV:1103.3481;%%
%``A Realistic Assessment of the Sensitivity of XENON10 and XENON100 to
%Light-Mass WIMPs,''
arXiv:1106.0653 .
%%CITATION = ARXIV:1106.0653;%%



\bibitem{otherdm}
See e.g.
D. N. Spergel and P. J. Steinhardt, Phys. Rev. Lett. 84, 3760 (2000) 
[astro-ph/9909386];
A. E. Faraggi and M. Pospelov, 
Astropart. Phys. 16, 451 (2002) [arXiv: hep-ph/0008223];
S. Mitra, Phys. Rev. D71 121302 (2005) [astro-ph/0409121];
J. L. Feng, H. Tu and H-B. Yu, JCAP 0810: 43 (2008) [arXiv: 0808.2318];
H. An, S-L. Chen, R. N. Mohapatra, S. Nussinov and Y. Zhang,
Phys. Rev. D82, 023533 (2010) [arXiv: 1004.3296];
N. Fornengo, P. Panci and M. Regis, arXiv: 1108.4661;
J-W. Cui, H-J. He, L-C. Lv and F-R. Yin, arXiv: 1110.6893.

\bibitem{collar}
J. I. Collar and F. T. Avignone III, Phys. Lett. B275, 181 (1992);
Phys.Rev. D47 5238 (1993); F. Hasenbalg {\it et al.},
Phys.Rev. D55, 7350 (1997).
 
\bibitem{arg}
D. E. Di Gregorio {\it et al.},
Nucl.Phys.Proc.Suppl. 48, 56 (1996). 

\bibitem{foot2008}
R. Foot, Phys. Rev. D78, 043529 (2008) [arXiv: 0804.4518].

\bibitem{he}
R. Foot and X-G. He, Phys. Lett. B267, 509 (1991). 

\bibitem{holdom}
B. Holdom, Phys. Lett. B166, 196 (1986).

\bibitem{footz}
R. Foot, Phys. Lett. B699, 230 (2011) [arXiv:1011.5078].


\bibitem{review}
R. Foot, Int. J. Mod. Phys. D13, 2161 (2004)
[astro-ph/0407623]; P. Ciarcelluti, Int. J. Mod. Phys. D19: 2151 (2010) [arXiv: 1102.5530].

\bibitem{some}
H. M. Hodges, Phys. Rev. D47, 456 (1993); 
Z. Berezhiani, D. Comelli and F. L. Villante,
Phys. Lett. B503, 362 (2001) [hep-ph/0008105];
L. Bento and Z. Berezhiani, Phys. Rev. Lett. 87, 231304 (2001)
[hep-ph/0107281]; 
A. Yu. Ignatiev and R. R. Volkas, Phys. Rev. D68, 023518 (2003)
[hep-ph/0304260];
R. Foot and R. R. Volkas, Phys. Rev. D68, 021304 (2003)
[hep-ph/0304261]; Phys. Rev. D69, 123510 (2004) [hep-ph/0402267];
Z. Berezhiani, P. Ciarcelluti, D. Comelli and F. L. Villante,
Int. J. Mod. Phys. D14, 107 (2005) [astro-ph/0312605];
P. Ciarcelluti, Int. J. Mod. Phys. D14, 187 (2005) [astro-ph/0409630];
Int. J. Mod. Phys. D14, 223 (2005) [astro-ph/0409633].
For pioneering work, see: S. I. Blinnikov and M. Yu. Khlopov, Sov. J. Nucl. Phys.
36, 472 (1981); Sov. Astron. 27, 371 (1983).

\bibitem{sph}
R. Foot and R. R. Volkas,
Phys. Rev. D70, 123508 (2004) [astro-ph/0407522].

\bibitem{footold1}
R. Foot, Phys. Rev. D69, 036001 (2004) [hep-ph/0308254].

\bibitem{footold}
R. Foot, Mod. Phys. Lett. A19,
1841 (2004) [astro-ph/0405362]; Phys. Rev. D74, 023514 (2006) [astro-ph/0510705].



%\bibitem{lab1}
%R. Foot, A. Yu. Ignatiev and R. R. Volkas, 
%Phys. Lett. B503, 355 (2001) [arXiv: astro-ph/0011156];
%R. Foot, Int. J. Mod. Phys. A19 3807 (2004) [astro-ph/0309330];
%R. Foot and Z. K. Silagadze, Int. J. Mod. Phys. D14, 143 (2005) [astro-ph/0404515];
%R. Foot, Phys. Lett. B699, 230 (2011) [arXiv:1011.5078].
%See also, S. Davidson, S. Hannestad and G. Raffelt, JHEP 5, 3
%(2000) [arXiv: hep-ph/0001179].

%\bibitem{mitra}
%R. Foot and S. Mitra, Astropart. Phys. 19, 739 (2003) [astro-ph/0211067]; 
%Phys. Lett. A315, 178 (2003) [cond-mat/0306561];
%Phys. Lett. B558, 9 (2003) [astro-ph/0301229].

\bibitem{p3}
P. Ciarcelluti and R. Foot, Phys. Lett. B690, 462 (2010) [arXiv:1003.0880].

\bibitem{slimit}
A.~Y.~Ignatiev and R.~R.~Volkas,
%``Geophysical constraints on mirror matter within the earth,''
Phys.\ Rev.\  D62, 023508 (2000) [arXiv:hep-ph/0005125].
%%CITATION = PHRVA,D62,023508;%%


\bibitem{earthmodel} 
A.~M.~Dziewonski and D.~L.~Anderson,
%``Preliminary Reference Earth Model,''
Phys.\ Earth Planet.\ Interiors 25, 297 (1981).
%%CITATION = PEPIA,25,297;%%


\bibitem{disf}
P. Ciarcelluti and R. Foot, 
Phys. Lett. B679, 278 (2009) [arXiv: 0809.4438].


\bibitem{de}
R. Bernabei {\it et al.}, (DAMA Collaboration),
Nuovo Cim. A112, 1541 (1999).

\bibitem{wong}
H. Wong, On behalf of the CDEX-Texono Collaboration, Talk at TAUP, Munich, September 2011.

\bibitem{sp}
R. Maruyama, on behalf of DM-ice collaboration, talk at TAUP, Sept. 2011.

%\bibitem{helm}
%R. H, Helm, Phys. Rev. 104, 1466 (1956). 

%\bibitem{smith}
%J. D. Lewin and P. F. Smith, Astropart. Phys. 6, 87 (1996).


\end{thebibliography}
\end{document}